# Dynamic maintenance policy for systems with repairable components subject to mutually dependent competing failure processes

Nooshin Yousefi [a*], David W. Coit [a,b], Zhu, Xiaoyan[c]
[a] *Department of Industrial & System Engineering, Rutgers University, Piscataway, NJ*, 08854
[b] *Tsinghua University, Beijing, China*
[c]*School of Economics and Management, University of Chinese Academy of Sciences, Beijing, China*
* *Corresponding author, [no.yousefi@rutgers.edu](mailto:no.yousefi@rutgers.edu), Tel: +1 304-276-3097*

**Abstract**:

In this paper, a repairable multi-component system is studied where all the components can be repaired individually within the system. The whole system is inspected at inspection intervals and the failed components are detected and replaced with a new one, while the other components continue functioning. Replacing components individually within the system makes their initial age to be different at each inspection time. Different initial age of all the components have affect on the system reliability and probability of failure and consequently the optimal inspection time, which is for the whole system not individual components. A dynamic maintenance policy is proposed to find the next inspection time based on the initial age of all the components. Two competing failure processes of degradation and a shock process are considered for each component. In our paper, there is a mutual dependency between the degradation process and shock process. Each incoming shock adds additional abrupt damages on the cumulative degradation path of all the components, and the shock arrival process is affected by the system degradation process. A realistic numerical example is presented to illustrate the proposed reliability and maintenance model.

*Keywords*: Individually repairable component, Inspection interval, Multi-component system, gamma process, mutually dependent competing failure process.

## 1. Introduction:

This paper is an extension of the previous studies for reliability analysis and maintenance of a multi-component system with individually repairable components. Each component is subject to multiple failure processes of degradation and random shock considering the mutual dependency of competing failure processes.

There have been several studies on dependencies between shock and degradation processes, which can be categorized into two groups of shock-degradation dependence and degradation-shock dependence models. In shock-degradation dependence model, each incoming shock can cause some damage on the degradation process which can be in different forms of adding an abrupt increase on the degradation path, increasing the hazard rate of sudden failure, and increasing the degradation rate. In degradation-shock dependence models, the intensity of shock arrival, the damage of each shock, or the hard failure threshold can be affected by the degradation process. However, many industrial systems and products can experience the mutual dependence of degradation and shock processes. Each incoming shock can change the degradation path and at the same times the current degradation level can affect the intensity of coming shock. In this paper, the mutual dependency of degradation and shock process is considered for a system with individually repairable components.

For systems with individually repairable components, each component can be replaced individually upon its failure. Previous research often considered the repairable systems as systems packaged and sealed together, so it is impossible to repair each component individually. However, for some industrial systems, repairing the individual components within the system is more cost effective than repairing the whole system upon failure of each component. At each inspection time, by replacing each failed component, the age or the degradation level of components within the system would be different; in other word, each component has its own age at the beginning of each inspection interval. Considering the initial age of all the components within the system, where each of them experience the mutual dependent competing failure processes is a challenge which is studied in this paper. A reliability model and dynamic inspection planning is proposed for a system with individually repairable components where each of them are subject to dependent competing failure processes considering the mutual dependency of degradation and shock processes. Each incoming shock causes an abrupt damage on the degradation path of each component. Moreover, the combination of all the components degradation level can affect the occurrence intensity of shock process.

From the maintenance perspective, there are different types of multi-component systems. The first type is a system with multiple components which each of them degrades individually but it is not practical or beneficial to maintain them individually which are studied in the previous research [1-3]. Examples of this kind of systems are cellphones, or micro-electro-mechanical system (MEMS) which are consist of different components which degrade individually but for maintenance perspective, they should be packaged together. The second type of systems are consisting of different components that degrade differently, but they can be maintained individually within the system. Wind turbines, railroad, some manufacturing systems such as hydraulic valve, etc., are examples of multi-component system which consists of different components which can be maintained individually. Moreover, in practice, there are mutual dependence between gradual deterioration of systems and the environmental shocks process. For example, for a hydraulic valve the arrival contamination locks will cause the abrupt wear debris or its component such as sleeves and spool, and the intensity of the contamination locks will increase with wear progressing and the number of contamination locks increasing so it can be conducted that the degradation process and shock process are mutually dependent.

Maintenance is a significant contributor to the total company's cost, so optimal maintenance policies in terms of cost, equipment downtime and quality should be identified [4].

The maintenance function is defined as a set of activities or tasks used to restore an item to a state in which it can perform its designated functions [5]. Industrial systems and products experience multiple failure modes which may be competing and dependent. Gradual degradation process and catastrophic failure process are the two common failure modes in many systems which can be referred to as dependent competing failure processes. The degradation process presents the level of system physical deterioration over time, and the soft failure occurs when the degradation level of a system or component reaches a predefined failure threshold. Hard failure is the second failure mode which occurs due to instantaneous stress caused by shock process.

The paper is organized as follows. Section 2 introduces related previous research on reliability and maintenance policies for complex systems. Section 3 analyzes the reliability of a repairable multi-component system subject to mutually dependent competing failure processes and maintenance model. A realistic numerical example is given in Section 4, and the conclusion is presented in Section 5.

The notation used in formulating the reliability and maintenance models is listed as follows:

| | | |
|---|---|---|
| $\alpha_i(t), \beta_i$ | = | shape and scale parameter for gamma degradation process for component $i$; |
| $N(t)$ | = | number of shock loads that have arrived by time $t$; |
| $\lambda_0$ | = | Initial intensity of random shocks; |
| $\lambda_j(t)$ | = | Intensity of random shocks at t when $j$ shocks have arrived |
| $D_i$ | = | threshold for catastrophic/hard failure of $i^{th}$ component; |
| $W_{ij}$ | = | size/magnitude of the $j^{th}$ shock load on the $i^{th}$ component; |
| $F_{Wi}(w)$ | = | cumulative distribution function (cdf) of $W_i$; |
| $H_i$ | = | soft failure threshold of the $i^{th}$ component; |
| $Y_{ij}$ | = | damage size contributing to soft failure of the $i^{th}$ component caused by the $j^{th}$ shock load; |
| $f_{Y_i}(y)$ | = | probability density function (pdf) of $Y_i$; |
| $f_{Y_i}^{<k>}(y)$ | = | pdf of the sum of $k$ independent and identically distributed (*i.i.d.*) $Y_i$ variables |
| $X_i(t)$ | = | degradation of the $i^{th}$ component at $t$; |
| $X_{Si}(t)$ | = | total degradation of the $i^{th}$ component at $t$ due to continuous degradation, shock damage and initial degradation at the beginning of the inspection interval; |
| $X_S(t)$ | = | surrogate measurement for the system degradation |
| $G_i(x_i,t)$ | = | cumulative distribution function (cdf) of $X_i(t)$; |
| $\eta$ | = | Facilitation factor |
| $\gamma$ | = | Dependence factor |
| $\tau$ | = | inspection interval; |
| $CR(\tau,\boldsymbol{u})$ | = | average cost rate function of the maintenance policy; |
| $E[\rho]$ | = | expected system downtime (the expected time from a system failure to the next inspection when the failure is detected); |
| $C_R$ | = | replacement cost per unit; |
| $C_I$ | = | cost associated with each inspection; |
| $C_\rho$ | = | penalty cost rate during downtime per unit of time; |

## 2. Background

Significant studies have been done on reliability analysis and maintenance optimization for single systems and multi-component systems subject to multiple failure processes. Kharoufeh et al. [6] developed a reliability model for a single unit system subject to Markovian deterioration. Zhu et al. [7] studied the reliability and maintenance of a single unit system subject to competing risks. Chatwattanasiri et al. [8] studied the reliability of multi-component systems with uncertain future conditions. Lin and Pham [9] studied the reliability of a system where its configuration changes dynamically. Sun et al [10] analyzed the reliability of a multi-component system considering a Markov model for the system structure. Tian and Liao [11] investigated condition-based maintenance policies of multi-component systems based on a proportional hazards model. Yousefi et al. [12] developed a new maintenance model for a multi-component system subject to dependent competing failure processes.

There have been several studies considering shock-degradation dependence models. Peng et al [13] developed a reliability model for a system experiencing the degradation process and shock process, when each incoming shock causes an abrupt damage on the system degradation path. Rafiee et al. [14] developed a reliability model for a multi-component system considering the changes on the degradation rate as the shock damages. Song et al [15] proposed a reliability model for a multi-component system subject to multiple competing failure processes considering different shock sets. Yousefi et al [16] studied the system reliability and dynamic inspection plan of a multi-component system subject to shock and degradation processes considering shock-degradation dependency. Zhu et al [17] studied the reliability and maintenance of a *k*-out-of-*n*:F data storage system considering the degradation and external shock arrival . Jiang et al [18] considered different damage zones for each incoming shock and developed a reliability model for a system based on this shock-degradation dependence model. Cha et al. [19] and Yang et al. [20] considered two damage types for each incoming shock, where it can cause some abrupt damage on the degradation process and increase the hazard rate of sudden failure at the same time. Mercier et al. [21] investigated the interactive relationship between the degradation and the shock damage, where each arriving shock increases of the degradation rate, and the fatality of a shock may differ between the components.

Degradation-shock dependence model is one of the topics have been researched recently. Caballé et al [22] modeled the shock process as a doubly stochastic Poisson process where its intensity depends on the cumulative degradation process. Zhu et al [23] developed the reliability model of a system when the hard failure threshold is dependent on the cumulative degradation process. Fan et al [24] investigated the reliability model of a system subject to dependent competing failure processes where the next incoming shock magnitude increases by the degradation process. Ye et al [25] studied the reliability of a system where the damage probability of each incoming shock depends on the remaining hazard in the degradation process.

In practice, systems can be subject to mutual dependency of degradation and shock process. Che et al [26] proposed a reliability model for a system subject to degradation and shock processes, considering mutually dependent shock and degradation processes, where each incoming shock has an abrupt damage on the cumulative degradation path and the current degradation level can change the occurrence intensity of shock process. Yousefi and Coit [27] proposed a reliability model for a system subject to mutually dependent competing failure processes; where the shock arrival rate is affected by the cumulative degradation path and each shock has double effects on the degradation process, (a) changing the degradation rate and (b) causing an additional abrupt damage on the degradation process.

A repairable system is the one that can be restored to its functioning state by implementing some repair processes. Scholars have been investigated several studies on reliability analysis and maintenance models of repairable systems. Peng et al [28] studied the reliability analysis of a repairable system with partially relevant failures. Yen et al [29] investigated the reliability analysis of a repairable system considering warm standby and working breakdown condition. Wang and Zhang [30] developed an algorithm to find the optimal maintenance inspection plan for a repairable system. Lin and Huang [31] proposed a preventive maintenance policy for a repairable system considering a nonhomogeneous Poisson process with a power law failure intensity to describe the system deterioration. In previous research, a repairable system was generally considered to be packaged and sealed together, and the whole system was repaired or replaced upon failure. However, for some multi-components system, it is not beneficial and cost effective to replace the whole system upon failure of any components. For systems with repairable components, each component can be repaired or replaced individually within the system. Therefore, at any inspection time, the health status of all of the components should be checked and the failed components are replaced/repaired at the beginning of the next inspection.

In this paper, a multi-component system with individually repairable components is considered where each component experiences two mutually dependent competing failure processes of degradation and shock processes. Each incoming shock causes incremental damage on the cumulative degradation of each components. Moreover, the occurrence intensity of each shock is affected by the current cumulative degradation of all the components within the systems and the number of shocks received by the system.

## 3. System reliability analysis

In a multi-component system experiencing two dependent competing failure processes of soft and hard failure, degradation of each component can be calculated by adding the cumulative shock damages to the pure degradation of each component[13]. By pure degradation, we mean the degradation without the shock damages and it is modeled by gamma process. At any inspection time, component *i* fails due to soft failure if its cumulative degradation is greater than the predefine threshold ($H_i$) and fails due to the hard failure if any shock magnitude is greater than the hard failure threshold ($D_i$). Figure 1 shows the soft and hard failure processes on any component of a system.

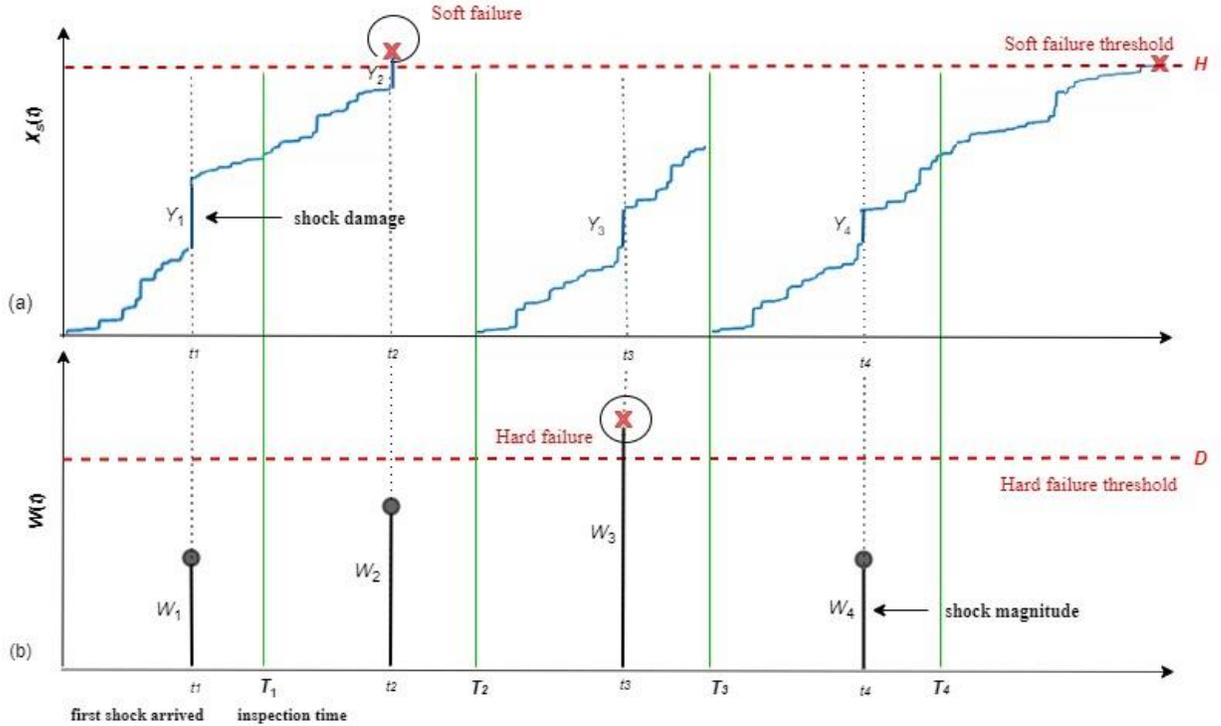

Figure 1 (a) Soft failure process and (b) Hard failure processes

For a multi-component system, each component degrades in the form of cumulative damage and a stochastic process can be used to model the degradation of each component. Abdul-Hameed [32] initially proposed the gamma process as a deterioration model. Van Noortwijk [33] reviewed the gamma process and its applications. A gamma process is used to model the degradation of each component $i$. In our new model, the gamma process, with linear shape parameter $\alpha_i(t) = \alpha t$ and scale parameter $\beta_i$ is a continuous time stochastic process with the following properties:

- It starts from 0 at time 0, i.e., $X_i(0) = 0$
- $X_i(t)$ has independent increment
- For $t>0$ and $s>0$, $X_i(t)-X_i(s) \sim \text{gamma}(\alpha_i(t)- \alpha_i(s), \beta_i)$.

In fact, the probability density function of degradation process for each component is given by:

$$g(x_i; \alpha_i(t-s), \beta_i) = \frac{\beta_i^{\alpha_i(t-s)} x_i^{\alpha_i(t-s)-1} \exp(-\beta_i x_i)}{\Gamma(\alpha_i(t-s))} \tag{1}$$

$\alpha_i(t)$ and $\beta_i$ are the shape parameter and scale parameter for component $i$. It is assumed that the damage shock on each component follows a normal distribution, $Y_{ij} \sim Normal(m_{Y_i}, S^2_{Y_i})$ where $Y_{ij}$ is the $j^{th}$ shock damage on the $i^{th}$ component. The total degradation of each component $i$ is the summation of its pure degradation process and damage caused by shocks.

It is also assumed that the shock magnitude is an *i.i.d* random variable which follows normal distribution $W_{ij} \sim Normal(\mu_{Wi}, \sigma^2_{Wi})$, where $W_{ij}$ is the $j^{th}$ shock magnitude for component $i$. Any other distribution could be used without loss of generality. Therefore, the probability of no hard failure for component $i$, due to a single shock magnitude of $j$, is as follow ($P_{NHi}$).

$$P_{NH_i} = P(W_{ij} < D_i) = F_{W_i}(D_i) = \Phi(\frac{D_i - \mu_{W_i}}{\sigma_{W_i}}) \tag{2}$$

### 3.1. *Reliability analysis of multi-component system with individually repairable components*

For a multi-component system with individually repairable components, upon observation of a failure at any inspection time, instead of replacing the whole system, the failed component should be replaced individually with a new one, but other components which are not failed, continue functioning until they fail. It is also assumed that all the components are inspected at a periodic time interval, that is an inspection interval, and no continuous monitoring is performed. Since we replace the failed component instead of the whole system, the ages of components at each inspection time are different from the others. In this paper, random variable $u_i$ are assumed as the initial age of component $i$ at the beginning of each inspection interval. Figure 2 shows the different age of all the components at each inspection interval.

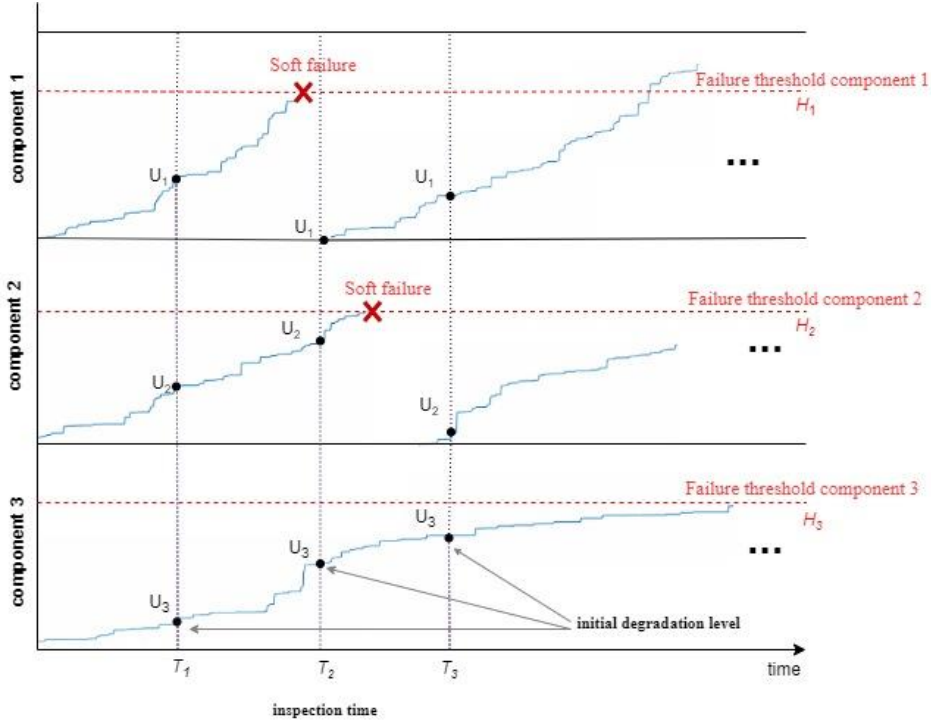

Figure 2 different initial ages among all the components of a system

The conditional reliability of component *i* at the beginning of an inspection interval when it is subject to mutual dependent competing failure processes can be calculated as follow; in other word, it is the probability of no soft failure nor hard failure by time *t*, after the previous inspection for component *i*.

$$R_i(t;u_i) = \sum_{m=0}^{\infty} \left[ \left( F_{W_i}(D_i) \right)^m P\left( X_i(t) + \sum_{j=1}^{N(t)} Y_{ij} + u_i < H_i \right) \Big| P(N(t) = m) \right] P(N(t) = m) \tag{3}$$

Where $P(N(t)=m)$ is the probability of having *m* shocks by time *t*. If the shock arrivals are independent of all the other arrivals in the process, it can be modeled as a Poisson process. So, the probability of having *m* shocks by time *t* can be calculated as follow

$$P(N(t) = m) = \frac{(\lambda t)^m e^{-\lambda t}}{m!} \tag{4}$$

In this paper, it is assumed that the components are configured as a series system, so, the system reliability for a multi-component system with individually repairable component subject to mutual dependent competing failure can be calculated as follow.

$$R(t;\mathbf{u}) = \sum_{m=0}^{\infty}\prod_{i=1}^{n}\left[P(W_{ij}<D_i)^m \times \int_0^{H_i-u_i} P\left(X_i(t)<H_i-y-u_i\right)f_{Y_i}^{<m>}(y)dy\right]P(N(t)=m) \quad (5)$$

$$= \sum_{m=0}^{\infty}\prod_{i=1}^{n}\left[(F_{W_i}(D_i))^m \int_0^{H_i-u_i} G(H_i-y-u_i;\alpha_i(t),\beta)f_{Y_i}^{\langle m\rangle}(y)\right]P(N(t)=m)$$

The probability of having *m* shocks by time *t* can be calculated using equation (4) when the arrivals occurs independently. However, in this research extension, the next shock arrival depends on the number of shocks that have already arrived by the system, so the Poisson process is not an appropriate method to model the shock arrival process. Therefore, the new shock arrival process is explained more in the next subsection.

### 3.2. *Reliability analysis of multi-component system subject to mutually dependent competing processes*

The cumulative degradation process of all components which is summation of initial degradation, pure degradation and damages from shocks affects the occurrence intensity of next shock arrival. Figure 3 shows the shock-degradation dependence and degradation-shock dependence in the proposed model.

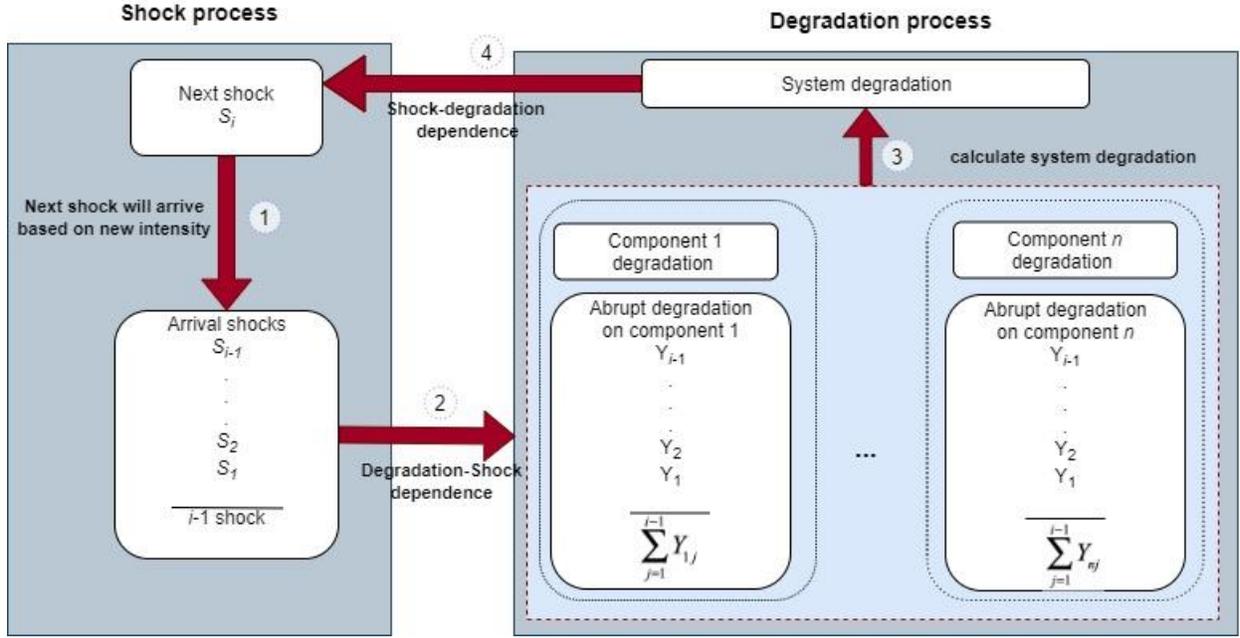

Figure 3. mutual dependence model for shock and degradation process

A Poisson process is a random process when each event is independent of all other events in the process. However, in this study the shock arrival depends on the number of shocks arriving to the system and the total degradation, so the Poisson process is not suitable for the shock arrival process. Che et al [26] showed that random shock process $\{N(t), t>0\}$ can be modeled as a facilitation model using intensity function $\lambda_i(t)$. As it is proved in [26] the probability of having $i$ shocks by time $t$ can be calculated as the following equation.

$$P_i(t) = \frac{(\eta^{-1} + (i-1))^i (\exp(\eta \Lambda_0(t) - 1)^i (\exp(-(1+\eta i)\Lambda_0(t))))}{i!} \qquad (6)$$

$$= \binom{\eta^{-1}+i-1}{i}(1-\exp(-\eta\Lambda_0(t)))^i(\exp(-\eta\Lambda_0(t)))^{\eta^{-1}}$$

Where $\Lambda_i(t) = \int_0^t \lambda_i(t)$ and the intensity function is $\lambda_i(t) = (1+\eta i)\lambda_0(t)$ which depends on the total degradation $\lambda_0(t) = \lambda_0 + \gamma X_S(t)$, where in Che et al [26] the total degradation referred to one component; however, in this paper, the shock process is for the system and $X_S(t)$ will be the sum of components degradation. $\gamma$ indicates the effect of the current degradation levels on the intensity and $\eta$ is the facilitation factor which shows the effect of abrupt degradation on the shock process. Considering the initial age of each component $i$ at the beginning of inspection intervals, the total degradation of each component $i$ is the summation of initial degradation of component $i$, $u_i$, and the pure degradation of component $i$ and all the shock damages on component $i$ by time $t$.

$$X_{S_i}(t) = X_i(t) + \sum_{j=1}^{N(t)} Y_{ij} + u_i.$$

The conditional probability of having no soft failure by time $t$ is ($P_{NS}$)

$$P_{NS_i}(t;u_i) = P(X_{S_i}(t) < H_i) = \sum_{m=0}^{\infty} P(X_i(t) + \sum_{j=1}^{N(t)} Y_{ij} + u_i < H_i) \times P(N(t)=m) \qquad (7)$$

The conditional survival probability of component $i$ by time $t$ can be calculated as follow:

$$R_i(t;u_i) = \sum_{m=0}^{\infty} \left[ (F_{W_i}(D_i))^m P\left(X_i(t) + \sum_{j=1}^{N(t)} Y_{ij} + u_i < H_i\right) \mid N(t)=m \right] P(N(t)=m) \qquad (8)$$

$$= \sum_{m=0}^{\infty} \left[ (F_{W_i}(D_i))^m \int_0^{H_i - u_i} G(H_i - y - u_i; \alpha_i(t), \beta) f_{Y_i}^{\langle m \rangle}(y) \right] P(N(t)=m)$$

Equation (6) is for single component and depends on degradation amount, but not for a system with $n$ components. To extend this model, a degradation measurement is needed for system total degradation to determine the shock intensity. Summation of all of the component degradation is now used as a surrogate measurement for the system degradation $X_S(t) = \sum_{i=1}^{n} X_{S_i}(t)$. Therefore, the probability that we have $m$ shocks by time $t$ from the beginning of the inspection for a system with series configurated components is as follow

$$P(N(t)=m) = \binom{\eta^{-1}+m-1}{m}(1-\exp(-\eta\Lambda_0(t)))^m (\exp(-\eta\Lambda_0(t)))^{\eta^{-1}}$$

$$= \binom{\eta^{-1}+m-1}{m}\left(1-\exp\left(-\eta \times \int_0^t \left(\lambda_0 + \gamma(\sum_{i=1}^n x_{S_i}(v))\right)dv\right)\right)^m \times \left(\exp\left(-\eta\int_0^t \left(\lambda_0 + \gamma(\sum_{i=1}^n x_{S_i}(v))\right)dv\right)\right)^{\eta^{-1}}$$

(10)

Where $x_{s_i}(v)$ is the actual degradation path realization for $0 \leq v \leq t$, and $\int_0^t x_{s_i}(v)dv$ is unique for each specific degradation path and approximated by the sum of $\Delta x_{s_i}(v)$ for $v$ from 0 to $t$.

To estimate the parameters of the proposed method, if a sufficient degradation dataset is available, Maximum Likelihood Estimation (MLE) method can be applied to estimate all the parameters. In MLE, the parameter values are found such that they maximize the likelihood of the process. The likelihood function can be formulated by calculating the probability density function of observed degradation level at each inspection time and multiplying them together. Based on the problem, different optimality search algorithm can be applied to find the optimal parameter values.

### 3.3. *Maintenance modeling*

In this paper, we proposed a dynamic inspection planning for a system with repairable components. Based on the equations in the previous subsections, it can be concluded that the system reliability and probability of failure are affected by the initial age of all the components within the system. So, in the proposed maintenance policy, each successive inspection interval should be found dynamically based on the initial age of all the components. For a system with multiple-components, that degrade differently, a preventive maintenance model should be found considering the age of all the components at the beginning of each interval. Figure 4 shows a system with three different repairable components. At the beginning of any inspection time, the initial age of all the components should be detected and the system reliability is calculated based on the initial age of all the components, and the next inspection time can be found subsequently.

In this paper, the only maintenance activity is assumed to be replacement. At any inspection time, component $i$ is detected as failed one, if its total degradation is greater than its failure threshold ($H_i$) or it has received any shock with magnitude of greater than its hard failure threshold ($D_i$). Therefore, component $i$ should be replaced with a new one, while the other components continue functioning. For a system with multiple-components, that degrade differently, the next inspection time should be found considering the age of all the components at the beginning of each interval.

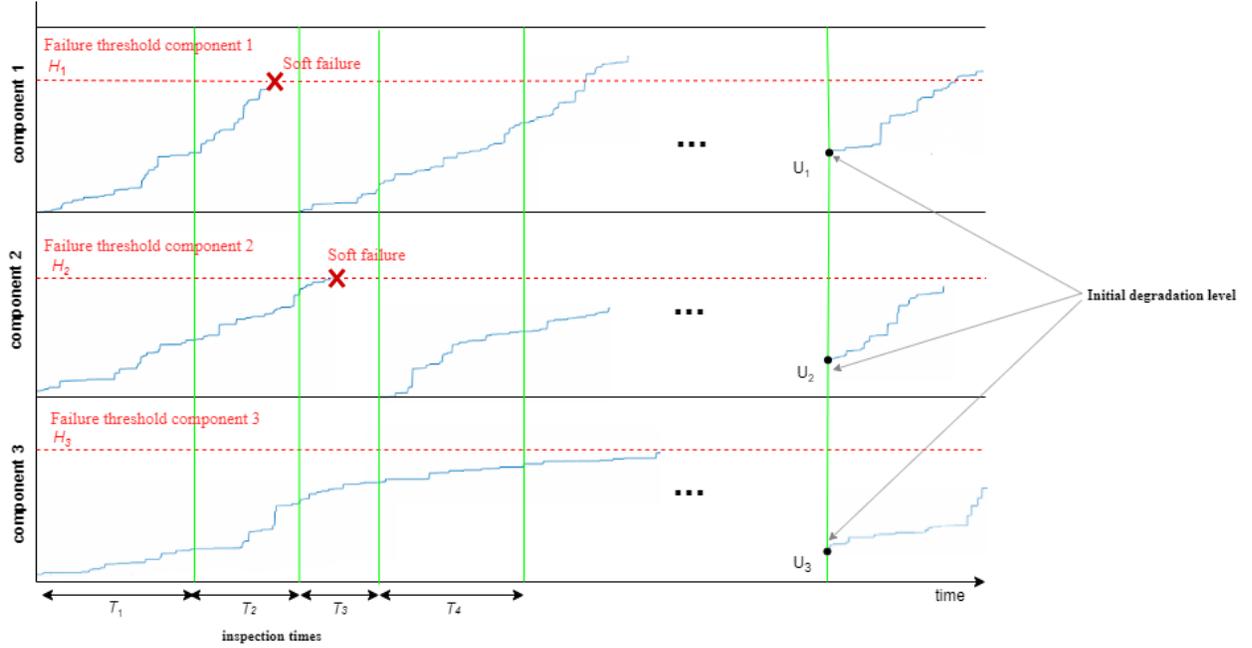

Figure 4 Dynamic inspection planning considering random initial age for each component [34]

In this paper, it is assumed that the whole system is inspected at each inspection time and there is no continuous monitoring. To find the optimal inspection interval for the next inspection, cost rate function is calculated and optimized dynamically.

$$CR(\tau;\mathbf{u}) = \frac{C_I + C_R(1-R(\tau;\mathbf{u})) + C_\rho E[\rho]}{\tau} \tag{11}$$

$C_I$ is the inspection cost, $C_R$ is the cost of replacement, $C_\rho$ is the downtime cost, $\tau$ is the inspection time and $E[\rho]$ is the expected of downtime. In this paper, the functions are evaluated numerically using Monte Carlo simulation with $10^5$ replications to find the reliability function, and cost rate function. The next optimal inspection interval at the beginning of the previous inspection time is found by numerical search to locate the minimum cost rate. The reliability is calculated using the observed total degradation of all the components. Appendix A shows the procedure of Monte Carlo simulation and optimization.

## 4. Numerical results

In this section, a realistic numerical example of a jet pipe servo-valve is considered. As the indispensable precision equipment in modern applications, the electrohydraulic servo-valves regulate the oil pressure, flux, and flow direction by controlling their opening size and on-off state to realize various motions of executive components [35]. The valve control hydraulic oil flows by the spool sliding in the sleeve [36]. However, the oil pollution can lead to the breakdown of leakage and the contamination lock (i.e., the clamping stagnation failure [37]). In this paper, the contamination lock is considered as hard failure and the wear between spool and sleeve is considered as soft failure. Table 1 shows the parameters for reliability analysis of this example. It is assumed that the inspection cost is $5, the downtime cost is $100, the replacement is $20.

Table 1 Parameters value

| Parameters | Description | Spool | Sleeve | Source |
|---|---|---|---|---|
| $H_i$ | Soft failure threshold | 5 mm | 6 mm | Fan et al. [38] |
| $D_i$ | Hard failure threshold | 40 N | 45 N | Che et al. [26] |
| $\alpha_i$ | Shape parameter of gamma process | 0.5 | 0.2 | Assumption |
| $\beta$ | Scale parameter of gamma process | 1.2 | 1.6 | Assumption |
| $\lambda_0$ | Initial intensity of random shock | $2.5 \times 10^{-5}$ | $2.5 \times 10^{-5}$ | Fan et al. [38] |
| $\eta$ | Facilitation factor | 0.2 | 0.2 | Che et al. [26] |
| $\gamma$ | Dependence factor | 0.001 | 0.001 | Fan et al. [38] |
| $W_{ij}$ | Friction caused by contamination lock | $W_{ij} \sim N(10, 5^2)$ N | $W_{ij} \sim N(14, 3^2)$ N | Assumption |
| $Y_{ij}$ | Wear increase by contamination lock | $Y_{ij} \sim N(0.5, 0.1^2)$ mm | $Y_{ij} \sim N(0.55, 0.1^2)$ mm | Che et al. [26] |

The system reliability and optimal maintenance policy is presented for two different models for a jet pipe servo-valve.

Model 1: The shock arrivals are independent of previous arrivals; in other word, shock process is modeled as homogeneous Poisson process and there is no degradation-shock dependency in this model.

Model 2: Both degradation-shock dependence model and shock-degradation model is considered as mutually dependent competing failure processes for all the components.

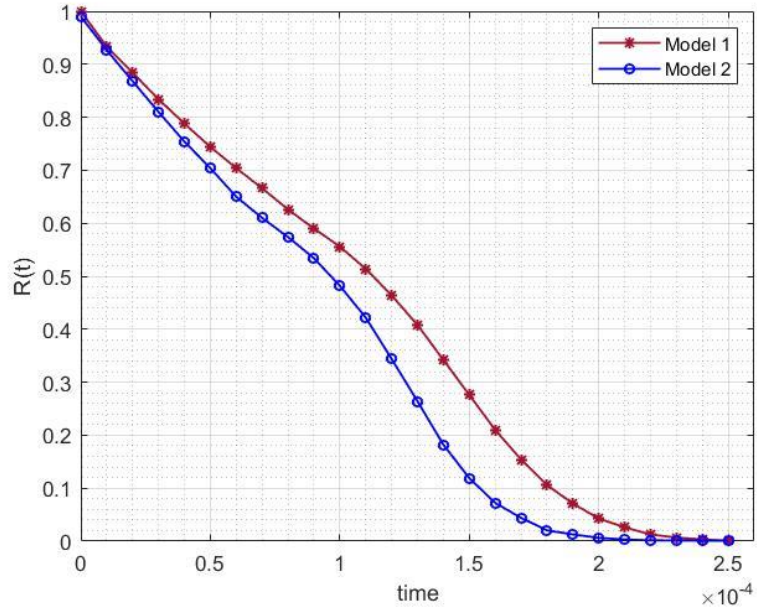

Figure 5 Reliability comparison of the proposed model and previous models

Figure 5 shows the system reliability for model 1 and model 2. When there is no degradation-shock dependency, the Poisson process is used to model the shock arrival process. As it is shown in Figure 5, the system reliability for model 1 with Poisson process as shock arrival model, is higher than the system reliability considering the mutual dependency of degradation and shock process. As the system degrades over time, the occurrence intensity of shocks arrival increases and the system receives more shocks, which subsequently increases the probability of failure, and decreases the system reliability at each point of time. Monte Carlo simulation with $10^5$ replications is used to calculate the system reliability for both models.

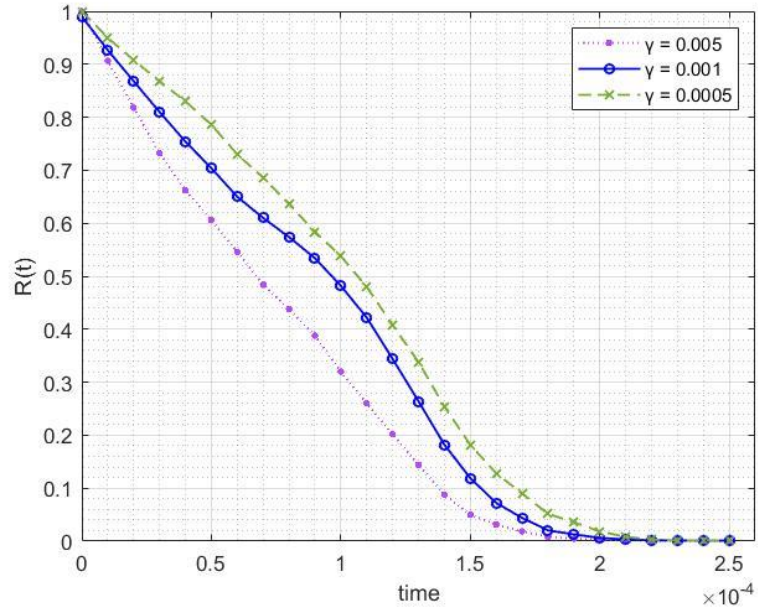

Figure 6 Sensitivity analysis of R(t) on γ

Figure 6 shows the sensitivity analysis of system reliability on degradation-shock dependence factor ($\gamma$). When $\gamma$ is small, there is less degradation-shock dependence, and high $\gamma$ makes the system receive more shocks which results faster deterioration, and subsequently lower system reliability, while there is almost no degradation-shock dependence when $\gamma$ is small, so the system reliability is higher.

To find the best inspection interval, we considered different scenarios, by assuming different values for initial age of each component and by running Monte Carlo simulation with $10^5$ replications, the cost function of each time interval is calculated, the time which is corresponded to the minimum cost function is the optimal inspection interval for that scenario. Figures 7 and 8 show the cost rate function of three different combinations of components' initial age for model 1 and model 2.

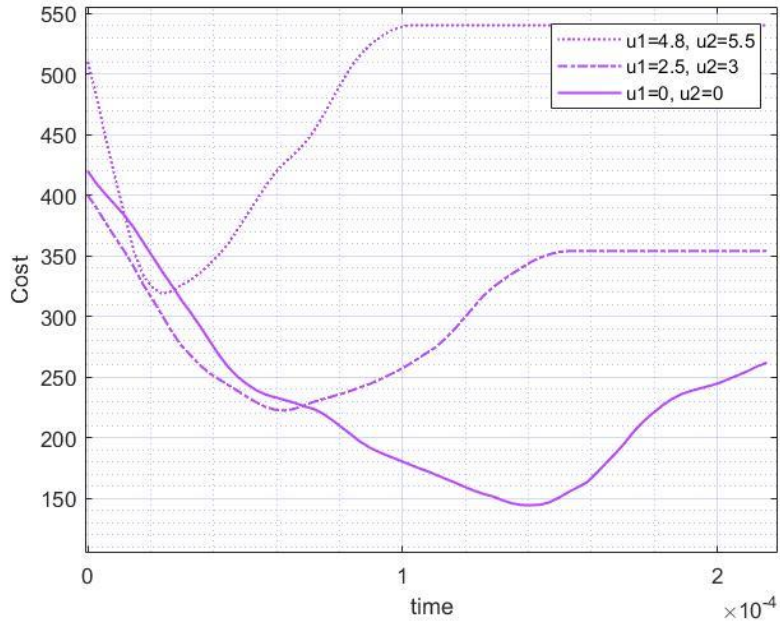

Figure 7 Cost rate for different combination of components' initial age for model 1

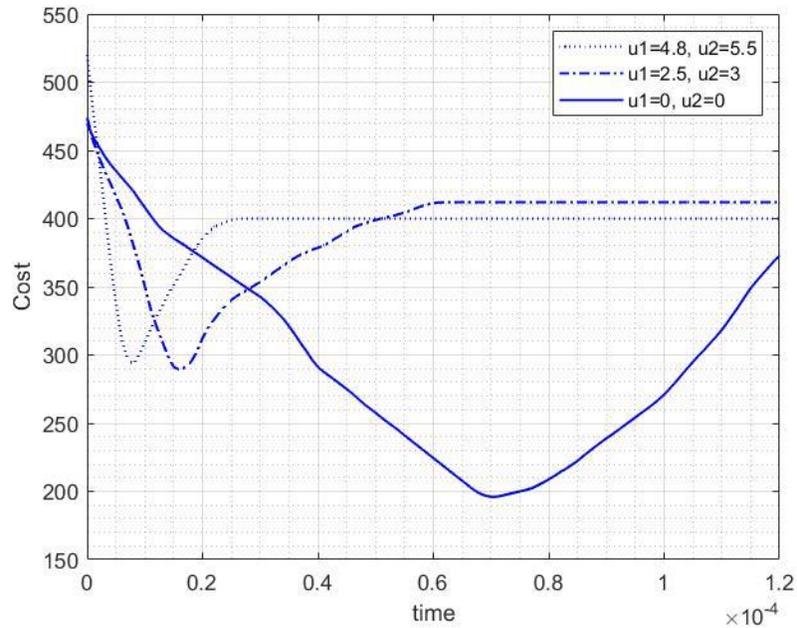

Figure 8 Cost rate for different combination of components' initial age for model 2

Table 2 also shows the optimal inspection interval for model 1 and model 2. When the initial age of all the components in the system are zero, the optimal inspection is higher than the scenario that the initial age of each components is close to its failure threshold. When the components are not new at the beginning of the interval, they are more prone to failure than the new components, and consequently they should be inspected in a short interval to avoid the failure and downtime cost.

Table 2 Optimal inspection interval for model1 and model2

| Scenarios number | Spool | Sleeve | Model 1, Optimal inspection interval ($\tau^*$) $\times 10^{-4}$ | Model 2, Optimal inspection interval ($\tau^*$) $\times 10^{-4}$ |
|---|---|---|---|---|
| 1 | 0 | 0 | 1.432 | 0.741 |
| 2 | 0.5 | 0.5 | 1.356 | 0.713 |
| 3 | 1 | 0.5 | 1.122 | 0.695 |
| 4 | 0.5 | 1 | 1.271 | 0.691 |
| 5 | 1 | 1 | 1.083 | 0.614 |
| 6 | 2 | 2 | 0.954 | 0.542 |
| 7 | 2.5 | 0.5 | 0.818 | 0.410 |
| 8 | 0.5 | 3 | 0.748 | 0.286 |
| 9 | 2.5 | 3 | 0.657 | 0.187 |
| 10 | 4 | 4 | 0.441 | 0.083 |
| 11 | 4.5 | 2 | 0.503 | 0.097 |
| 12 | 1 | 5.5 | 0.212 | 0.051 |
| 13 | 0.5 | 5.8 | 0.091 | 0.039 |
| 14 | 4.8 | 0.5 | 0.086 | 0.034 |
| 20 | 4.8 | 5.5 | 0.023 | 0.009 |

As it is shown in Figure 9 and Table 2, comparing the optimal inspection interval for model 1 and model 2 for the same scenarios, it can be concluded that the system of model 2 has higher probability of failure than model 1. The mutually dependency between shock process and degradation process, make the system in model 2 degrades faster; hence, to avoid the system failure and downtime cost the system should be inspect in a shorter interval than model 1.

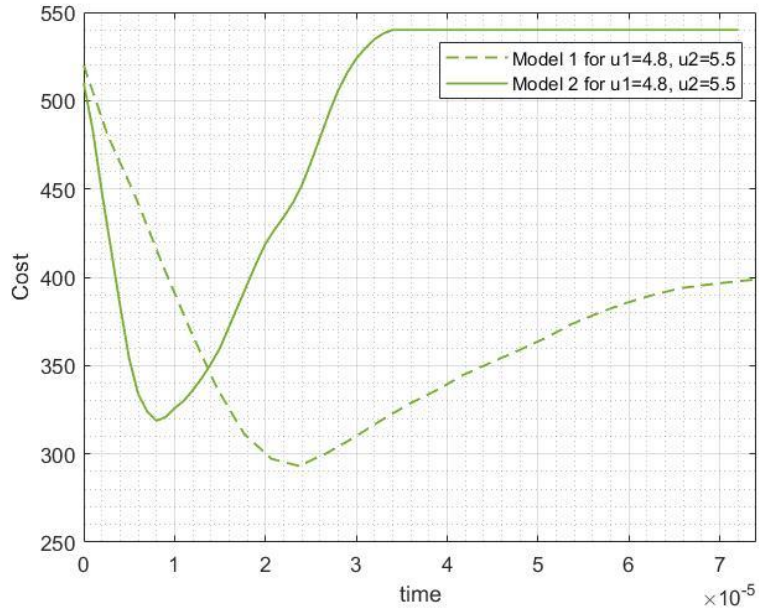

Figure 9 Comparison of cost function of model 1 and model 2 for one scenario

Figure 10 shows the reliability comparison of four different cases similar to other related models. By setting different values of the parameters we can calculate the reliability of four different cases of (1) no dependency between shock and degradation process, with $Y=0$, $\eta=0$ and $\gamma=0$ for both components (2) shock-degradation dependency with $Y_{ij} \sim Normal(\mu_{Y_i}, \sigma_{Y_i}^2)$ for component $i$ and $\eta=0$ and $\gamma=0$ for both components (3) degradation-shock dependency with $Y=0$, $\eta=0$ and $\gamma=0.001$ for both components (4) both degradation and shock process are mutually dependent with $Y_{ij} \sim Normal(\mu_{Y_i}, \sigma_{Y_i}^2)$ for component $i$ and $\eta=0.2$ and $\gamma=0.001$ for both components.

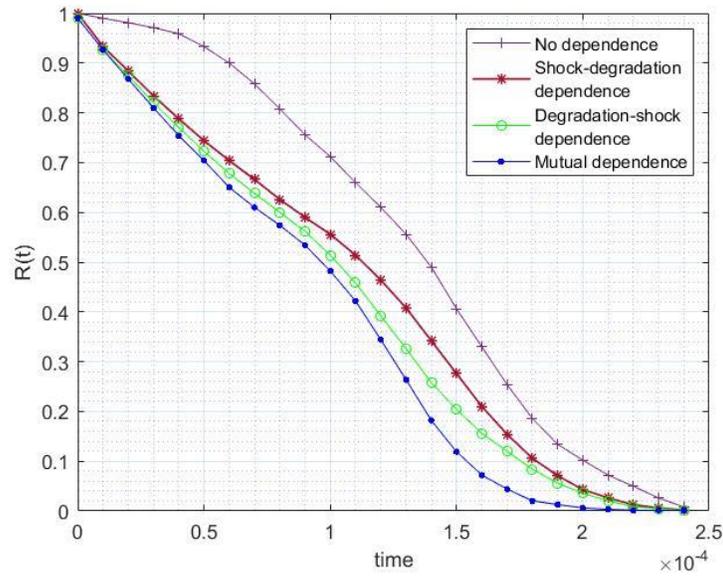

Figure 10 Reliability of four cases (1) no dependence (2) shock-degradation dependence (3) degradation-shock dependence (4) mutual dependence

While insights from the example are applicable for other systems with similar failure behavior, a detailed analysis should still be conducted. In this paper, an example of a jet pipe servo-valve is used to show how the proposed method can be applied to calculate the system reliability of a multi-component system experiencing mutually dependent competing failure processes. However, the proposed method can be used for calculating the system reliability of any system experiencing multiple failure processes where they are mutually dependent such as wind turbines, or rail roads which consists of multiple components where each of them experiencing degradation processes and random shock process like, weight stress, weather shocks, etc.

## 5. Conclusion

In this paper, a new reliability model and maintenance policy is developed for a multi-component system with individually repairable components, where each component is subject to two competing failure process of soft and hard failure. The mutual dependence of degradation and shock process is considered for such system. Each incoming shock has a damage on all the components which is considered an abrupt jump on the degradation path of all the components within the system. The cumulative degradation of all the components cause some changes on the occurrence intensity of shock arrivals. If the components are more deteriorated, they are more prone to receive shocks. The shock-degradation dependence and degradation-shock dependence models for a multi-component system with individually repairable components are considered in this paper. A new reliability model is developed for such system, and consequently a dynamic maintenance policy is proposed considering the age of all the components at the beginning of the inspection intervals. The different age of all the components have effect on the system reliability and probability of failure, so it is suggested that the next inspection interval should be found dynamically based on the initial age of components. Finding the next inspection interval considering the initial age of the components provides a cost-effective maintenance plan rather than the fixed inspection interval. A numerical example shows the performance of proposed reliability and dynamic maintenance policy. In this paper, a system with series configuration is considered while the dynamic inspection plan for other configuration can be evaluated in the future studies. However, in some cases all the incoming shocks will not have damage on the degradation level and subsequently on the intensity of the shock arrival, while in this paper, it is assumed that all the incoming shocks are fatal and have damage on the degradation, different types of shocks such as fatal shocks and non-fatal shocks can be considered in the extension of this paper in the future. Moreover, In this paper, only replacement is considered as the maintenance action, while there are different type of maintenance which can be applied such as imperfect or minimal repair which can be considered in the future study. Different shock patterns can also be considered for each component.

**Appendix A**

The procedure of Monte Carlo simulation and finding the optimal inspection interval numerically is as follow based on the algorithm proposed in [26]

- **Set** $t, u_1, u_2, \ldots, u_n, M$ (the maximum number of shocks), and $N$ (the number of replications)
- **For** $j = 1:N$,

    Calculate $x(t)$ using Equation (1) for all the components,
    Calculate probability of no hard failure using equation (2)

- **For** $m = 0:M$

    Sample $Y_i$ from $f_Y^{<m>}(y)$

    Calculate total degradation for each component using $X_{S_i}(t) = X_i(t) + \sum_{j=1}^{m} Y_{ij} + u_i$ and

    $$X_S(t) = \sum_{i=1}^{n} X_{S_i}(t)$$

    Calculate probability of having $m$ shocks by time $t$ using equation (10)

- **End For**

    Calculate $R_i(t; u_i) = \sum_{m=0}^{M} \left[ (F_{W_i}(D_i))^m \int_0^{H_i - u_i} G(H_i - y - u_i; \alpha_i(t), \beta) f_{Y_i}^{\langle m \rangle}(y) \right] P(N(t) = m)$ for

    component $i$

    Calculate the system reliability by $R(t) = \frac{1}{N} \sum_{j=1}^{N} R_j(t)$

- **For** $\tau = 0.1 \times 10^6 : 1$

    Calculate the cost rate using Equation (11) and find the optimal inspection time which has lowest cost rate.

---